\documentstyle[12pt,epsf]{article}
\catcode`\@=11
\def\@date{}
\def\figuretype{BMP}

\def\diag{\mathop{\hbox{\,\rm diag}}\nolimits}

\def\re{\mathop{\hbox{\,\rm Re}}\nolimits}
\def\im{\mathop{\hbox{\,\rm Im}}\nolimits}

\def\hr{\hbox{\bf R}}

\def\@{@}

\def\section{\@startsection {section}{1}{\z@}{-3.5ex plus-1ex minus
    -.2ex}{2.3ex plus.2ex}{\reset@font\large\bf}}
\def\subsection{\@startsection{subsection}{2}{\z@}{-3.25ex plus-1ex
    minus-.2ex}{1.5ex plus.2ex}{\reset@font\bf}}
\def\subsubsection{\@startsection{subsubsection}{3}{\z@}{-3.25ex plus
 -1ex minus-.2ex}{1.5ex plus.2ex}{\reset@font\small\bf}}
\def\@afterheading{\global\@nobreaktrue\everypar{\if@nobreak
   \global\@nobreakfalse\clubpenalty \@M
   \else \clubpenalty \@clubpenalty\everypar{}\fi}}

\def\FN@{\futurelet\next}
\def\DN@{\def\next@}
\def\DNii@{\def\nextii@}
\def\RIfM@{\relax\ifmmode}
\def\RIfMIfI@{\relax\ifmmode\ifinner}
\def\setboxz@h{\setbox\z@\hbox}
\def\wdz@{\wd\z@}
\def\boxz@{\box\z@}
\def\setbox@ne{\setbox\@ne}
\def\wd@ne{\wd\@ne}
\def\Invalid@#1{\def#1{\Err@{\Invalid@@\string#1}}}
\def\Invalid@@{Invalid use of }
\def\eat@#1{}
\def\Let@{\relax\iffalse{\fi\let\\=\cr\iffalse}\fi}
\def\vspace@{\def\vspace##1{\crcr\noalign{\vskip##1\relax}}}
\newif\ifinany@
\newif\ifinalign@
\newif\ifingather@
\def\strut@{\copy\strutbox@}
\newbox\strutbox@
\setbox\strutbox@\hbox{\vrule height8\p@ depth3\p@ width\z@}
\def\topaligned{\null\,\vtop\aligned@}
\def\botaligned{\null\,\vbox\aligned@}
\def\aligned{\null\,\vcenter\aligned@}
\def\aligned@{\bgroup\vspace@\Let@
 \ifinany@\else\openup\jot\fi\ialign
 \bgroup\hfil\strut@$\m@th\displaystyle{##}$&
 $\m@th\displaystyle{{}##}$\hfil\crcr}
\def\endaligned{\crcr\egroup\egroup}

\def\alignedat#1{\null\,\vcenter\bgroup\doat@{#1}\vspace@\Let@
 \ifinany@\else\openup\jot\fi\ialign\bgroup\span\preamble@@\crcr}
\newcount\atcount@
\def\doat@#1{\toks@{\hfil\strut@$\m@th
 \displaystyle{\the\hashtoks@}$&$\m@th\displaystyle
 {{}\the\hashtoks@}$\hfil}
 \atcount@#1\relax\advance\atcount@\m@ne                                    
 \loop\ifnum\atcount@>\z@\toks@=\expandafter{\the\toks@&\hfil$\m@th
 \displaystyle{\the\hashtoks@}$&$\m@th
 \displaystyle{{}\the\hashtoks@}$\hfil}\advance
  \atcount@\m@ne\repeat                                                     
 \xdef\preamble@{\the\toks@}\xdef\preamble@@{\preamble@}}

\def\gathered{\null\,\vcenter\bgroup\vspace@\Let@
 \ifinany@\else\openup\jot\fi\ialign
 \bgroup\hfil\strut@$\m@th\displaystyle{##}$\hfil\crcr}
\def\endgathered{\crcr\egroup\egroup}
\newif\iftagsleft@
\def\TagsOnLeft{\global\tagsleft@true}
\def\TagsOnRight{\global\tagsleft@false}
\TagsOnRight
\newif\ifmathtags@
\def\TagsAsMath{\global\mathtags@true}
\def\TagsAsText{\global\mathtags@false}
\TagsAsText
\def\tagform@#1{\hbox{\rm(\ignorespaces#1\unskip)}}
\def\thetag{\leavevmode\tagform@}
\def\tag#1$${\iftagsleft@\leqno\else\eqno\fi                                
 \maketag@#1\maketag@                                                       
 $$}                                                                        
\def\maketag@{\FN@\maketag@@}
\def\maketag@@{\ifx\next"\expandafter\maketag@@@\else\expandafter\maketag@@@@
 \fi}
\def\maketag@@@"#1"#2\maketag@{\hbox{\rm#1}}                                
\def\maketag@@@@#1\maketag@{\ifmathtags@\tagform@{$\m@th#1$}\else
 \tagform@{#1}\fi}
\interdisplaylinepenalty\@M
\def\allowdisplaybreaks{\RIfMIfI@
 \onlydmatherr@\allowdisplaybreaks\else
 \interdisplaylinepenalty\z@\fi\else\onlydmatherr@\allowdisplaybreaks\fi}
\Invalid@\allowdisplaybreak
\Invalid@\displaybreak
\Invalid@\intertext
\def\allowdisplaybreak@{\def\allowdisplaybreak{\crcr\noalign{\allowbreak}}}
\def\displaybreak@{\def\displaybreak{\crcr\noalign{\break}}}
\def\intertext@{\def\intertext##1{\crcr\noalign{\vskip\belowdisplayskip
 \vbox{\normalbaselines\noindent##1}\vskip\abovedisplayskip}}}
\newskip\centering@
\centering@\z@ plus\@m\p@
\def\align{\relax\ifingather@\DN@{\csname align (in
  \string\gather)\endcsname}\else
 \ifmmode\ifinner\DN@{\onlydmatherr@\align}\else
  \let\next@\align@\fi
 \else\DN@{\onlydmatherr@\align}\fi\fi\next@}
\newhelp\andhelp@
{An extra & here is so disastrous that you should probably exit^^J
and fix things up.}
\newif\iftag@
\newcount\and@
\def\align@{\inalign@true\inany@true
 \vspace@\allowdisplaybreak@\displaybreak@\intertext@
 \def\tag{\global\tag@true\ifnum\and@=\z@\DN@{&&}\else
          \DN@{&}\fi\next@}%
 \iftagsleft@\DN@{\csname align \endcsname}\else
  \DN@{\csname align \space\endcsname}\fi\next@}
\def\Tag@{\iftag@\else\errhelp\andhelp@\err@{Extra & on this line}\fi}
\newdimen\lwidth@
\newdimen\rwidth@
\newdimen\maxlwidth@
\newdimen\maxrwidth@
\newdimen\totwidth@
\def\measure@#1\endalign{\lwidth@\z@\rwidth@\z@\maxlwidth@\z@\maxrwidth@\z@
 \global\and@\z@                                                            
 \setbox@ne\vbox                                                            
  {\everycr{\noalign{\global\tag@false\global\and@\z@}}\Let@                
  \halign{\setboxz@h{$\m@th\displaystyle{\@lign##}$}
   \global\lwidth@\wdz@                                                     
   \ifdim\lwidth@>\maxlwidth@\global\maxlwidth@\lwidth@\fi                  
   \global\advance\and@\@ne                                                 
   &\setboxz@h{$\m@th\displaystyle{{}\@lign##}$}\global\rwidth@\wdz@        
   \ifdim\rwidth@>\maxrwidth@\global\maxrwidth@\rwidth@\fi                  
   \global\advance\and@\@ne                                                
   &\Tag@
   \eat@{##}\crcr#1\crcr}}
 \totwidth@\maxlwidth@\advance\totwidth@\maxrwidth@}                       
\def\displ@y@{\global\dt@ptrue\openup\jot
 \everycr{\noalign{\global\tag@false\global\and@\z@\ifdt@p\global\dt@pfalse
 \vskip-\lineskiplimit\vskip\normallineskiplimit\else
 \penalty\interdisplaylinepenalty\fi}}}
\def\black@#1{\noalign{\ifdim#1>\displaywidth
 \dimen@\prevdepth\nointerlineskip                                          
 \vskip-\ht\strutbox@\vskip-\dp\strutbox@                                   
 \vbox{\noindent\hbox to#1{\strut@\hfill}}
 \prevdepth\dimen@                                                          
 \fi}}
\expandafter\def\csname align \space\endcsname#1\endalign
 {\measure@#1\endalign\global\and@\z@                                       
 \ifingather@\everycr{\noalign{\global\and@\z@}}\else\displ@y@\fi           
 \Let@\tabskip\centering@                                                   
 \halign to\displaywidth
  {\hfil\strut@\setboxz@h{$\m@th\displaystyle{\@lign##}$}
  \global\lwidth@\wdz@\boxz@\global\advance\and@\@ne                        
  \tabskip\z@skip                                                           
  &\setboxz@h{$\m@th\displaystyle{{}\@lign##}$}
  \global\rwidth@\wdz@\boxz@\hfill\global\advance\and@\@ne                  
  \tabskip\centering@                                                       
  &\setboxz@h{\@lign\strut@\maketag@##\maketag@}
  \dimen@\displaywidth\advance\dimen@-\totwidth@
  \divide\dimen@\tw@\advance\dimen@\maxrwidth@\advance\dimen@-\rwidth@     
  \ifdim\dimen@<\tw@\wdz@\llap{\vtop{\normalbaselines\null\boxz@}}
  \else\llap{\boxz@}\fi                                                    
  \tabskip\z@skip                                                          
  \crcr#1\crcr                                                             
  \black@\totwidth@}}                                                      
\newdimen\lineht@
\expandafter\def\csname align \endcsname#1\endalign{\measure@#1\endalign
 \global\and@\z@
 \ifdim\totwidth@>\displaywidth\let\displaywidth@\totwidth@\else
  \let\displaywidth@\displaywidth\fi                                        
 \ifingather@\everycr{\noalign{\global\and@\z@}}\else\displ@y@\fi
 \Let@\tabskip\centering@\halign to\displaywidth
  {\hfil\strut@\setboxz@h{$\m@th\displaystyle{\@lign##}$}%
  \global\lwidth@\wdz@\global\lineht@\ht\z@                                 
  \boxz@\global\advance\and@\@ne
  \tabskip\z@skip&\setboxz@h{$\m@th\displaystyle{{}\@lign##}$}%
  \global\rwidth@\wdz@\ifdim\ht\z@>\lineht@\global\lineht@\ht\z@\fi         
  \boxz@\hfil\global\advance\and@\@ne
  \tabskip\centering@&\kern-\displaywidth@                                  
  \setboxz@h{\@lign\strut@\maketag@##\maketag@}%
  \dimen@\displaywidth\advance\dimen@-\totwidth@
  \divide\dimen@\tw@\advance\dimen@\maxlwidth@\advance\dimen@-\lwidth@
  \ifdim\dimen@<\tw@\wdz@
   \rlap{\vbox{\normalbaselines\boxz@\vbox to\lineht@{}}}\else
   \rlap{\boxz@}\fi
  \tabskip\displaywidth@\crcr#1\crcr\black@\totwidth@}}
\expandafter\def\csname align (in \string\gather)\endcsname
  #1\endalign{\vcenter{\align@#1\endalign}}
\Invalid@\endalign
\newif\ifxat@
\def\alignat{\RIfMIfI@\DN@{\onlydmatherr@\alignat}\else
 \DN@{\csname alignat \endcsname}\fi\else
 \DN@{\onlydmatherr@\alignat}\fi\next@}
\newif\ifmeasuring@
\newbox\savealignat@
\expandafter\def\csname alignat \endcsname#1#2\endalignat                   
 {\inany@true\xat@false
 \def\tag{\global\tag@true\count@#1\relax\multiply\count@\tw@
  \xdef\tag@{}\loop\ifnum\count@>\and@\xdef\tag@{&\tag@}\advance\count@\m@ne
  \repeat\tag@}%
 \vspace@\allowdisplaybreak@\displaybreak@\intertext@
 \displ@y@\measuring@true                                                   
 \setbox\savealignat@\hbox{$\m@th\displaystyle\Let@
  \attag@{#1}
  \vbox{\halign{\span\preamble@@\crcr#2\crcr}}$}%
 \measuring@false                                                           
 \Let@\attag@{#1}
 \tabskip\centering@\halign to\displaywidth
  {\span\preamble@@\crcr#2\crcr                                             
  \black@{\wd\savealignat@}}}                                               
\Invalid@\endalignat

\def\matrix{\,\vcenter\bgroup\Let@\vspace@
    \normalbaselines
  \m@th\ialign\bgroup\hfil$##$\hfil&&\quad\hfil$##$\hfil\crcr
    \mathstrut\crcr\noalign{\kern-\baselineskip}}
\def\endmatrix{\crcr\mathstrut\crcr\noalign{\kern-\baselineskip}\egroup
                \egroup\,}
\def\pmatrix{\left(\matrix}
\def\endpmatrix{\endmatrix\right)}

\newdimen\spreadmlines@
\def\format{\crcr\egroup\iffalse{\fi\ifnum`}=0 \fi\format@}
\newtoks\hashtoks@
\hashtoks@{#}
\def\format@#1\\{\def\preamble@{#1}%
 \def\l{$\m@th\the\hashtoks@$\hfil}%
 \def\c{\hfil$\m@th\the\hashtoks@$\hfil}%
 \def\r{\hfil$\m@th\the\hashtoks@$}%
 \edef\Preamble@{\preamble@}\ifnum`{=0 \fi\iffalse}\fi
 \ialign\bgroup\span\Preamble@\crcr}
\def\cases{\bgroup\spreadmlines@\jot\left\{\,\matrix\format\l&\quad\l\\}
\def\endcases{\endmatrix\right.\egroup}

\def\wd@ne{\wd\@ne}
\def\setbox@ne{\setbox\@ne}
\def\binrel@#1{\setboxz@h{\thinmuskip0mu
  \medmuskip\m@ne mu\thickmuskip\@ne mu$#1\m@th$}%
 \setbox@ne\hbox{\thinmuskip0mu\medmuskip\m@ne mu\thickmuskip
  \@ne mu${}#1{}\m@th$}%
 \setbox\tw@\hbox{\hskip\wd@ne\hskip-\wdz@}}
\def\overset#1\to#2{\binrel@{#2}\ifdim\wd\tw@<\z@
 \mathbin{\mathop{\kern\z@#2}\limits^{#1}}\else\ifdim\wd\tw@>\z@
 \mathrel{\mathop{\kern\z@#2}\limits^{#1}}\else
 {\mathop{\kern\z@#2}\limits^{#1}}{}\fi\fi}
\def\underset#1\to#2{\binrel@{#2}\ifdim\wd\tw@<\z@
 \mathbin{\mathop{\kern\z@#2}\limits_{#1}}\else\ifdim\wd\tw@>\z@
 \mathrel{\mathop{\kern\z@#2}\limits_{#1}}\else
 {\mathop{\kern\z@#2}\limits_{#1}}{}\fi\fi}

\def\setboxz@h{\setbox\z@\hbox}
\def\wdz@{\wd\z@}
\def\boxz@{\box\z@}
\newif\ifmsbmloaded@
\def\widetilde#1{
\ifmsbmloaded@
  \setboxz@h{$\m@th#1$}\ifdim\wdz@>\tw@ em\mathaccent"0\msbfam@5D{#1}\else
  \mathaccent"0365{#1}\fi
 \else\mathaccent"0365{#1}\fi}

\mathchardef\varGamma="0100
\mathchardef\varDelta="0101
\mathchardef\varTheta="0102
\mathchardef\varLambda="0103
\mathchardef\varXi="0104
\mathchardef\varPi="0105
\mathchardef\varSigma="0106
\mathchardef\varUpsilon="0107
\mathchardef\varPhi="0108
\mathchardef\varPsi="0109
\mathchardef\varOmega="010A

\def\allowmathbreak{\relax\ifmmode\ifinner\allowbreak\else
  \nonmathaerr@\allowmathbreak\fi\else\nonmathberr@\allowmathbreak\fi}

\def\theoremfont{\def\@thmfont{\rm}}
\@addtoreset{definition}{section}

\newtheorem{theorem}{\bf Theorem}

\newtheorem{proposition}{\bf Proposition}

\def\D{\dsize}

\def\bigskip{\vskip12pt}
\def\medskip{\vskip6pt}
\def\smallskip{\vskip2pt}

\def\voidtoken{}

\def\refby{}
\def\refpaper{}
\def\refjour{}
\def\refvol{}
\def\refpage{}
\def\refpages{}
\def\refyr{}
\def\refbook{}
\def\refpubl{}
\def\refpubladdr{}

\def\bibref{
  \global\def\refby{}
  \global\def\refpaper{}
  \global\def\refjour{}
  \global\def\refvol{}
  \global\def\refpage{}
  \global\def\refpages{}
  \global\def\refyr{}
  \global\def\refbook{}
  \global\def\refpubl{}
  \global\def\refpubladdr{}
  }
\def\by#1{\global\def\refby{#1}}
\def\paper#1{\global\def\refpaper{#1}}
\def\jour#1{\global\def\refjour{#1}}
\def\vol#1{\global\def\refvol{#1}}
\def\page#1{\global\def\refpage{#1}}
\def\pages#1{\global\def\refpage{#1}}
\def\yr#1{\global\def\refyr{#1}}
\def\book#1{\global\def\refbook{#1}}
\def\publ#1{\global\def\refpubl{#1}}
\def\publaddr#1{\global\def\refpubladdr{#1}}
\def\endbibref{{
\ifx\refby\voidtoken \else \refby\fi
\ifx\refpaper\voidtoken \else , {\it\refpaper\/}\fi
\ifx\refjour\voidtoken \else , \refjour\fi
\ifx\refbook\voidtoken \else , \refbook\fi
\ifx\refpubl\voidtoken \else , \refpubl\fi
\ifx\refpubladdr\voidtoken \else , \refpubladdr\fi
\ifx\refvol\voidtoken \else \ \refvol\fi
\ifx\refyr\voidtoken \else \ (\refyr)\fi
\ifx\refpage\voidtoken \else , \refpage\fi.\vskip2pt}}

\newif\iff@rstzcite\f@rstzcitetrue
\def\zcite{\@tempswafalse\@zcitex[]}
\def\@zcitex[#1]#2{\if@filesw\immediate\write\@auxout{\string\citation{#2}}\fi
  \let\@citea\@empty
  \@zcite{\@for\@citeb:=#2\do
    {\@citea\def\@citea{\raise.75ex\hbox{\scriptsize,}\penalty\@m}%
     \def\@tempa##1##2\@nil{\edef\@citeb{\if##1\space##2\else##1##2\fi}}%
     \expandafter\@tempa\@citeb\@nil
     \@ifundefined{b@\@citeb}{{\reset@font\bf ? }\@warning
       {Citation `\@citeb' on page \thepage \space undefined}}
     \hbox{{}\raise0.75ex\hbox{\scriptsize\iff@rstzcite [\hskip-0.05em \fi
     \csname b@\@citeb\endcsname}}\f@rstzcitefalse}}{#1}
     \hskip-0.3em\hbox{{}\raise0.75ex\hbox{\scriptsize]}}\f@rstzcitetrue}
\def\@zcite#1{#1} 

\def\maketitle{\par
 \begingroup
   \if@twocolumn
     \twocolumn[\@maketitle]%
     \else 
     \global\@topnum\z@
     \@maketitle \fi
     \@thanks
 \endgroup
 \setcounter{footnote}{0}%
 \let\maketitle\relax
 \let\@maketitle\relax
 \gdef\@thanks{}\gdef\@author{}\gdef\@title{}\let\thanks\relax}

\def\@maketitle{
 \null
 \vskip 2em
 \begin{center}%
  {\LARGE \@title \par}%
  \vskip 1.5em
  {\large
   \lineskip .5em
   \begin{tabular}[t]{c}\@author
   \end{tabular}\par}%
  \vskip 1em
  {\large \@date}%
 \end{center}%
 \par
 \vskip 1.5em}

\catcode`\@=\active

\def\figuretype{EPS}

\def\odotH{\ \raise 4pt\hbox{\accent23}\hskip-10pt\hbox{$H$}}
\def\D{\displaystyle}
\def\HH{\hbox to 0.1pt{\phantom{$\D\frac AB$}\hss}}

\begin{document}
\pagestyle{plain}
\baselineskip=12pt

\title{Localized Solitons of Hyperbolic su(N) AKNS System}

\author{\small Zixiang Zhou%
\thanks{Institute of Mathematics and Laboratory
of Mathematics for Nonlinear Sciences, Fudan University\hfill\break
\ \hbox to 12pt{} Supported by Chinese National Research Project
``Nonlinear Science'' and Chinese National Science Foundation for
Youth.}\\
\small Institute of Mathematics, Fudan University, Shanghai
200433, China\\
\small E-mail: zxzhou\@guomai.sh.cn} 

\maketitle

\begin{abstract}
Using the nonlinear constraint and Darboux transformation
methods, the $(m_1,\cdots,m_N)$ localized solitons of the
hyperbolic su(N) AKNS system are constructed. Here ``hyperbolic
su(N)'' means that the first part of the Lax pair is
$\varPsi_y=J\varPsi_x+U(x,y,t)\varPsi$ where $J$ is constant real
diagonal and $U^*=-U$. When different solitons move in different
velocities, each component $U_{ij}$ of the solution $U$ has at
most $m_im_j$ peaks as $t\to\infty$. This corresponds to the
$(M,N)$ solitons for the DSI equation. When all the solitons move
in the same velocity, $U_{ij}$ still has at most $m_im_j$ peaks
if the phase differences are large enough.
\end{abstract}

\section{Introduction}\label{sec:intro}

Since the discovery of localized solitons for the DSI
equation,\zcite{bib:BoitiBT} the $(M,N)$ solitons,\zcite{bib:FS}
especially the $(N,N)$ solitons, are discussed in various ways,
such as the inverse scattering method,\zcite{bib:BLP, bib:FSung,
bib:HMM, bib:Pemp, bib:Santini}, the binary Darboux
transformation method,\zcite{bib:MS} the Hirota 
method etc.\zcite{bib:HH} On the other hand, the nonlinear
constraint method\zcite{bib:Cao} has been developed for 1+2
dimensional problems since the work of \cite{bib:Kono, bib:CL}.
Using the Darboux transformation in higher
dimensions\zcite{bib:Gu, bib:GZN}, the solutions can be obtained
explicitly. In \cite{bib:ZhouNwave}, the non-localized solutions
of the N-wave equation were derived in this way. In
\cite{bib:Zhou2n}, the localized soliton solutions 
of the hyperbolic su(N) AKNS system were discussed by nonlinear
constraint method which transforms the original problem to a
$2N\times 2N$ linear system which separates all the variables. It
shows that, if the velocities of the solitons are all different,
there are Darboux transformations of $r$th degree such that the
asymptotic solution has at most $r^2$ peaks. However, the
velocities are not arbitrary. For example, the velocities of
``1-soliton'' should be along the $y$-axis. In \cite{bib:Li}, a
localized one-soliton solution of the \hbox{DSI\hskip-2pt
I\hskip-2ptI} equation was obtained by nonlinear constraint
method which transforms the original problem to an $(N+1)\times
(N+1)$ linear system which separates all the variables.

In this paper, we show that under the nonlinear constraint, there
is a way to construct Darboux transformation so that the number
of peaks and the values of velocities have more freedom. This
construction is valid for all the lower ordered equations in
the hyperbolic su(N) AKNS system, including the DSI equation, the
N-wave equation etc. For the DSI equation, this corresponds to the
$(M,N)$ solitons.  

The main conclusions are: (1) There are $m=(m_1,\cdots,m_N)$
solitons $U^{[m]}$ which are localized (tend to $0$ as
$(x,y)\to\infty$); (2) If different solitons move in different
velocities (characterized by an algebraic condition
(\ref{eq:diffv}) and hold for the DSI equation), then the
component $U_{jk}^{[m]}$ of $U^{[m]}$ has at most $m_jm_k$ peaks
as $t\to\infty$; (3) If all the solitons move in the same
velocity (hold for the N-wave equation), then $U_{jk}^{[m]}$ also
has at most $m_jm_k$ peaks as the phase differences are large
enough.  

\section{The system and its nonlinear constraints}

Here we consider the hyperbolic su(N) AKNS system
\begin{equation}
   \aligned
   &\varPsi_y=J\varPsi_x+U(x,y,t)\varPsi\\
   &\varPsi_t=\sum_{j=0}^n V_j(x,y,t)\partial^{n-j}\varPsi
   \endaligned
   \label{eq:LP}
\end{equation}
where
$\partial=\partial/\partial x$, $J=\diag(J_1,\cdots,J_N)$ is a
real constant diagonal $N\times N$ matrix with 
mutually different entries. $U(x,y,t)$ is off-diagonal with
$U^*=-U$. In this case, we call (\ref{eq:LP}) a hyperbolic su(N)
AKNS system, since $J$ is real and $U\in su(N)$. 

The integrability conditions of (\ref{eq:LP}) are
\begin{eqnarray}
   &&[J,V_{j+1}^A]=V_{j,y}^A-JV_{j,x}^A-[U,V_j]^A
    +\sum_{k=0}^{j-1} C_{n-k}^{n-j}(V_k\partial^{j-k}U)^A
    \label{eq:orgA}\\
   &&V_{j,y}^D-JV_{j,x}^D=[U,V_j^A]^D
    -\sum_{k=0}^{j-1} C_{n-k}^{n-j}(V_k\partial^{j-k}U)^D
    \label{eq:orgD}\\
   &&U_t=V_{n,y}^A-JV_{n,x}^A-[U,V_n]^A
    +\sum_{k=0}^{n-1} (V_k\partial^{n-k}U)^A \label{eq:orgevol}
\end{eqnarray}
where the superscripts $D$ and $A$ refer to the diagonal and
off-diagonal parts of a matrix. As mentioned in
\cite{bib:Zhou2n}, (\ref{eq:orgD}) and (\ref{eq:orgevol}) give a
system of nonlinear PDEs of $U$ and all the diagonal parts of
$V_j$'s, where $V_j$'s are determined by (\ref{eq:orgA})
respectively. Usually, only $U$ is physically significant.

There is a connection of (\ref{eq:LP}) with the linear system
\begin{equation}
   \aligned
   &\varPhi_x=\pmatrix i\lambda I &iF\\ iF^* &0 \endpmatrix,\qquad
    \varPhi_y=\pmatrix i\lambda J+U &iJF\\ iF^*J &0 \endpmatrix\\
   &\varPhi_t=\pmatrix W &X\\ Y &Z \endpmatrix\varPhi
    =\sum_{j=0}^n \pmatrix W_j &X_j\\ -X_j^* &Z_j \endpmatrix 
    \lambda^{n-j}\varPhi
   \endaligned
   \label{eq:LP2}
\end{equation}
where $F$, $W_j$, $X_j$, $Z_j$ are $N\times K$, $N\times N$,
$N\times K$, $K\times K$ matrices respectively ($K\ge 1$)
and satisfy $W_j^*=-W_j$, $Z_j^*=-Z_j$. This is a slight
generalization of the linear system in \cite{bib:Zhou2n}, where
$K$ should be $N$.

The integrability conditions of (\ref{eq:LP2}) include 
\begin{equation}
   \aligned
   &F_y=JF_x+UF\\
   &iF_t=X_{n,x}+iW_nF-iFZ_n
   \endaligned
   \label{eq:redLP}
\end{equation}
\begin{equation}
   \aligned
   &iX_{j+1}=X_{j,x}+iW_jF-iFZ_j\\
   &W_{j,x}=-iFX_j^*-iX_jF^*\\
   &Z_{j,x}=iF^*X_j+iX_j^*F\\
   &i[J,W_{j+1}]=W_{j,y}-[U,W_j]+iJFX_j^*+iX_jF^*J\\
   &Z_{j,y}=iF^*JX_j+iX_j^*JF
   \endaligned
   \label{eq:redrecu}
\end{equation}
\begin{equation}
   U_x=[J,FF^*]
   \label{eq:rednlc}
\end{equation}
\begin{equation}
   U_t=W_{n,y}-[U,W_n]+iJFX_n^*+iX_nF^*J
   \label{eq:redevol}
\end{equation}

For $U=0$, $F=0$, (\ref{eq:redrecu}) implies that
$W_j(\lambda)=i\varOmega_j(t)$, $X_j=0$, $Z_j=iZ_j^0(t)$ where 
$\varOmega_j(t)$'s are real diagonal matrices and $Z_j^0(t)$'s
are real matrices. 

When $Z_j^0(t)=\zeta_j(t)I_K$ ($I_K$ is the
$K\times K$ identity matrix) where $\zeta_j(t)$ is a real function of
$t$, (\ref{eq:redLP}) is just the Lax pair (\ref{eq:LP}) for
$n=1,2,3$. (\ref{eq:redrecu}) and (\ref{eq:redevol}) give the
recursion relations to determine $W_j$, $X_j$, $Z_j$, together
with the evolution equations corresponding to
(\ref{eq:orgA})--(\ref{eq:orgevol}), which are the integrability 
conditions of (\ref{eq:LP}). (\ref{eq:rednlc}) gives a nonlinear
constraint between $U$ and $F$. 

This system includes the DSI equation and the N-wave equation as
special cases.

In order to consider the asymptotic behavior of the solution $U$,
here we suppose $\varOmega_j$ is independent of $t$ and
$\zeta_j=0$. Denote $\varOmega=\sum_{j=0}^n
\varOmega_j\lambda^{n-j}$ and write
$\varOmega=\diag(\omega_1,\cdots,\omega_N)$. 

We need the following symbols and simple conclusions.

If $M_1$, $M_2$ are $j\times k$ matrices, we write $M_1\doteq
M_2$ if there is a nondegenerate diagonal $k\times k$ matrix $A$
such that $M_2=M_1A$.  


If $L$ is a diagonal matrix, then $M_1\doteq M_2$ and 
$\det M_1\ne 0$ imply $M_1LM_1^{-1}=M_2LM_2^{-1}$.

Let 
\begin{equation}
   M=\pmatrix a &-v^*/\bar a\\ v &I_K \endpmatrix
\end{equation}
where $v\ne 0$ is an $K\times 1$ vector, $a\ne 0$ is a number. Let
\begin{equation}
   \varLambda=\pmatrix \lambda_0 &\\ &\bar\lambda_0I_K \endpmatrix.
\end{equation}
Then we have
\begin{equation}
   M^{-1}=\frac 1\varDelta \pmatrix \bar a &v^*\\ 
    -\bar av &\varDelta I_K-vv^* \endpmatrix
   \label{eq:M-1}
\end{equation}
\begin{equation}
   M\varLambda M^{-1}=\frac 1\varDelta \pmatrix 
    \bar\lambda_0\varDelta+(\lambda_0-\bar\lambda_0)|a|^2 
    &(\lambda_0-\bar\lambda_0)av^*\\
    (\lambda_0-\bar\lambda_0)\bar av 
    &\bar\lambda_0\varDelta I_K+(\lambda_0-\bar\lambda_0)vv^* \endpmatrix
   \label{eq:SS}
\end{equation}
where $\varDelta=v^*v+|a|^2$. Moreover,
\begin{equation}
   \lim_{a\to\infty}M\varLambda M^{-1}=\pmatrix 
    \lambda_0 &\\ &\bar\lambda_0I_K \endpmatrix
   \label{eq:SSinfty}
\end{equation}
\begin{equation}
   \lim_{a\to 0}M\varLambda M^{-1}=\pmatrix
    \bar\lambda_0 &\\ 
    &\D\bar\lambda_0I_K+(\lambda_0-\bar\lambda_0)\frac{vv^*}{v^*v}
    \endpmatrix
   \label{eq:SS0}
\end{equation}

For an $(N+K)\times(N+K)$ matrix $M$, denote $M_{B_N}$ be an
$N\times N$ matrix containing the first $N$ columns and $N$ rows
of $M$. 

\section{Darboux transformation and soliton solutions}\label{eq:DT}

Now we construct the solutions from $U=0$, $F=0$. A procedure to
construct the Darboux transformation was proposed in
\cite{bib:Zhou2n}. 

Let $\lambda_\alpha$ ($\alpha=1,2,\cdots,r$) be $r$ complex
numbers such that $\lambda_\alpha\ne\lambda_\beta$ for
$\alpha\ne\beta$ and $\lambda_\alpha\ne\bar\lambda_\beta$ for all
$\alpha$, $\beta$. Take
\begin{equation} 
   \varLambda_\alpha=\diag(
   \underbrace{\lambda_\alpha,\cdots,\lambda_\alpha}_N,
   \underbrace{\bar\lambda_\alpha,\cdots,\bar\lambda_\alpha}_K)
\end{equation}
\begin{equation}
   H_\alpha=\pmatrix \exp(Q_\alpha(s)) &-\exp(-Q_\alpha(s)^*)C_\alpha^*\\
    C_\alpha &I_K \endpmatrix
   \label{eq:defH}
\end{equation}
where
\begin{equation}
   Q_\alpha(s)=\pmatrix a_{\alpha 1}s+b_{\alpha 1} &&\\ &\ddots &\\ 
    &&a_{\alpha N}s+b_{\alpha N} \endpmatrix
   \label{eq:defQ}
\end{equation}
$a_{\alpha j}$, $b_{\alpha j}$ are constants,
\begin{equation}
   C_\alpha=(0,\cdots,0,
    \underset{\scriptstyle l_\alpha}\to{\kappa_\alpha},0,\cdots,0)
\end{equation}
$\kappa_\alpha$ is a constant $K\times 1$ nonzero vector which
appears at the $l_\alpha$'s column of $C_\alpha$.

Let
\begin{equation}
   \rho_\alpha=\re(a_{\alpha,l_\alpha})\qquad 
   \phi_\alpha=\im(a_{\alpha,l_\alpha})\qquad 
   \pi_\alpha=b_{\alpha,l_\alpha}
\end{equation}

The Darboux matrices for such $\{\varLambda_\alpha, H_\alpha\}$ can
be constructed as follows. Let
\begin{equation}
   \aligned
   &G^{(1)}(\lambda)=\lambda-H_1\varLambda_1H_1^{-1}\qquad
    H_\alpha^{(1)}=G^{(1)}(\lambda_\alpha)H_\alpha\quad
    (\alpha=2,3,\cdots,r)\\
   &G^{(2)}(\lambda)=\lambda-H_2^{(1)}\varLambda_2H_2^{(1)-1}\qquad
    H_\alpha^{(2)}=G^{(2)}(\lambda_\alpha)H_\alpha^{(1)}\quad
    (\alpha=3,4,\cdots,r)\\
   &\cdots\\
   &G^{(r)}(\lambda)=\lambda-H_r^{(r-1)}\varLambda_rH_r^{(r-1)-1}
   \endaligned
\end{equation}
\begin{equation}
   G(\lambda)=G^{(r)}(\lambda)G^{(r-1)}(\lambda)\cdots G^{(1)}(\lambda)
   \label{eq:defG}
\end{equation}
then $G(\lambda)$ is a polynomial of $\lambda$ of degree $r$. The
permutability\zcite{bib:stnbook} implies that if
$(\varLambda_\alpha,H_\alpha)$ and $(\varLambda_\beta,H_\beta)$
are interchanged, $G(\lambda)$ is invariant.  

Let
\begin{equation}
   m_j=\#\{\,\alpha\,|\,1\le\alpha\le r,\,l_\alpha=j\,\}\qquad
   m=(m_1,\cdots,m_N)
\end{equation}
then $m_1+\cdots+m_N=r$.

Suppose
\begin{equation}
   G(\lambda)=\lambda^r-G_1\lambda^{r-1}+\cdots+(-1)^rG_r
\end{equation}
denote
\begin{equation}
   U^{[m]}=i[J,(G_1)_{B_N}]
\end{equation}

For the problem (\ref{eq:LP2}) with $U=F=0$, we have
$Q_\alpha=i\lambda_\alpha(x+Jy)+i\varOmega(\lambda_\alpha)t$, where
$s$ can be $t$ or other parameters. The matrix $G(\lambda)$ given
by (\ref{eq:defG}) is a Darboux matrix, that is, for any solution
$\varPhi$ of (\ref{eq:LP2}), $G\varPhi$ satisfies (\ref{eq:LP2})
with certain $\widetilde U$, $\widetilde F$, $\widetilde W_j$,
$\widetilde X_j$, $\widetilde Z_j$ replacing $U$, $F$, $W_j$,
$X_j$, $Z_j$. $U^{[m]}$ is actually the derived solution of
(\ref{eq:redevol}). Owing to the choice of $H_\alpha$ in
(\ref{eq:defH}), $U^{[m]}$ is globally defined. Here we first
consider the generalized $Q_\alpha(s)$ in (\ref{eq:defQ}) so that the
localization, asymptotic properties etc.\ can be treated uniformly.

\begin{proposition}\label{theprop}
(1)~If there is at most one $\alpha$ $(1\le\alpha\le r)$ such that
$\rho_{\alpha}=0$, then $\lim_{s\to\infty}U^{[m]}=0$.\par  
(2)~If $\rho_{\alpha}=0$, $\rho_{\beta}=0$ $(\alpha\ne
\beta)$, and $\rho_{\gamma}\ne 0$ for all $\gamma\ne
\alpha,\beta$, then:\par
\noindent(i)~when $l_\alpha=l_\beta$, $\lim_{s\to\infty}U^{[m]}=0$;\par
\noindent(ii)~when $l_\alpha\ne l_\beta$, 
\begin{equation}
   \lim_{s\to\infty}U_{ab}^{[m]}=0\quad
   \hbox{for }(a,b)\ne(l_\alpha,l_\beta)
\end{equation}
and as $s\to\infty$,
\begin{equation}
   U_{l_\alpha,l_\beta}^{[m]}
    \sim\frac{B_{\alpha\beta}\exp(i\im(\pi_\alpha-\pi_\beta)
     +i(\phi_\beta-\phi_\alpha)s)}
    {A_{\alpha\beta}\cosh(\re(\pi_\alpha+\pi_\beta)
     -\delta_{\alpha\beta}^{(1)})
    +\cosh(\re(\pi_\alpha-\pi_\beta)
     -\delta_{\alpha\beta}^{(2)})}
\end{equation}
where $A_{\alpha\beta}$, $\delta_{\alpha\beta}^{(1)}$,
$\delta_{\alpha\beta}^{(2)}$, are real constants,
$A_{\alpha\beta}>0$, and $B_{\alpha\beta}$ is a complex constant.
Moreover, if $K=1$, then $B_{\alpha\beta}\ne 0$ if and only if
$\kappa_\alpha\ne 0$ and $\kappa_\beta\ne 0$. 
\end{proposition}

\begin{demo}
First suppose
$\rho_\alpha\ne 0$. By (\ref{eq:SSinfty}) and (\ref{eq:SS0}),
\begin{equation}
   \lim_{\rho_\alpha s\to\pm\infty}H_\alpha\varLambda_\alpha H_\alpha^{-1}
    =S_\alpha^{\pm\infty}
\end{equation}
where
\begin{equation}
   \aligned
   &S_\alpha^{+\infty}=\pmatrix \lambda_\alpha I_N &\\ 
    &\bar\lambda_\alpha I_K \endpmatrix\\
   &S_\alpha^{-\infty}=\pmatrix 
    \lambda_\alpha I_N+(\bar\lambda_\alpha-\lambda_\alpha)
     E_{l_\alpha l_\alpha} &\\
    &\D\bar\lambda_\alpha I_K+(\lambda_\alpha-\bar\lambda_\alpha)
     \frac{\kappa_\alpha\kappa_\alpha^*}{\kappa_\alpha^*\kappa_\alpha}
   \endpmatrix
   \endaligned
\end{equation}
$E_{jk}$ is an $N\times N$ matrix whose $(j,k)$th entry is $1$ and the
rest entries are zero.

For $\beta\ne \alpha$,
\begin{equation}
   (\lambda_\beta-S_\alpha^{\pm\infty})H_\beta\doteq\pmatrix
   \exp(Q_\beta(s)) &-\exp(-Q_\beta(s)^*)\widetilde C_\beta^{\pm*}\HH\\
    \widetilde C_\beta^{\pm} &I_K \endpmatrix
\end{equation}
where
\begin{equation}
   \aligned
   &\widetilde C_\beta^\pm
    =(0,\cdots,0,\underset{\scriptstyle l_\beta}\to
    {\widetilde\kappa_\beta^\pm},0,\cdots,0)\\
   &\widetilde\kappa_\beta^+
    =\frac{\lambda_\beta-\bar\lambda_\alpha}
     {\lambda_\beta-\lambda_\alpha}\kappa_\beta\\
   &\widetilde\kappa_\beta^-
    =\left\{\matrix
     \D\frac{\lambda_\beta-\bar\lambda_\alpha}
      {\lambda_\beta-\lambda_\alpha}\kappa_\beta
      -\frac{\lambda_\alpha-\bar\lambda_\alpha}
      {\lambda_\beta-\lambda_\alpha}
      \frac{\kappa_\alpha^*\kappa_\beta}
      {\kappa_\alpha^*\kappa_\alpha}\kappa_\alpha\hfill
     &\hbox{if }l_\beta\ne l_\alpha\hfill\\
     \D\kappa_\beta
      -\frac{\lambda_\alpha-\bar\lambda_\alpha}
      {\lambda_\beta-\bar\lambda_\alpha}
      \frac{\kappa_\alpha^*\kappa_\beta}
      {\kappa_\alpha^*\kappa_\alpha}\kappa_\alpha\hfill
     &\hbox{if }l_\beta=l_\alpha\hfill\endmatrix\right.
   \endaligned
\end{equation}
Therefore, if $\rho_\alpha\ne 0$, the action of the limit Darboux matrix
$\lambda-S_\alpha^{\pm\infty}$ on $H_\beta$ ($\beta\ne \alpha$)
does not change the form of $H_\beta$ , but only changes the
constant vector $\kappa_\beta$. 

If $K=1$, then $\kappa_\beta^{\pm *}\kappa_\gamma^{\pm}\ne 0$
implies $\widetilde\kappa_\beta^{\pm
*}\widetilde\kappa_\gamma^{\pm}\ne 0$. When $K>1$, this does not 
hold in general.

Now suppose $\rho_\alpha=0$. Without loss of generality, suppose
$l_\alpha=1$. Then
\begin{equation}
   H_\alpha\doteq\pmatrix 
   \exp(\pi_{\alpha}) &&&&-\exp(-\bar \pi_{\alpha})\kappa_\alpha^*\\
   &1 &&&0\\ &&\ddots &&\vdots\\ &&&1 &0\\
   \kappa_\alpha &0 &\cdots &0 &I_K \endpmatrix
\end{equation}
By (\ref{eq:SS}),
\begin{equation}
   \aligned
   &H_\alpha\varLambda_\alpha H_\alpha^{-1}=\frac 1\varDelta\cdot\\
   &\cdot\pmatrix
   \bar\lambda_\alpha\varDelta+(\lambda_\alpha-\bar\lambda_\alpha)
   \exp(\pi_{\alpha}+\bar \pi_{\alpha}) 
   &&&&(\lambda_\alpha-\bar\lambda_\alpha)\exp(\pi_{\alpha})
    \kappa_\alpha^*\\
   &\lambda_\alpha &&&0\\ &&\ddots &&\vdots\\ &&&\lambda_\alpha &0\\
   (\lambda_\alpha-\bar\lambda_\alpha)\exp(\bar \pi_{\alpha})
    \kappa_\alpha &0 &\cdots &0 
   &\bar\lambda_\alpha\varDelta I_K
    +(\lambda_\alpha-\bar\lambda_\alpha)\kappa_\alpha\kappa_\alpha^*
   \endpmatrix
   \endaligned
   \label{eq:0S}
\end{equation}
where
$\varDelta=\exp(\pi_\alpha+\bar\pi_\alpha)
+\kappa_\alpha^*\kappa_\alpha$.

Part (1) of the proposition is derived as follows. Owing to the
permutability of Darboux transformations, we can suppose
$\rho_1\ne 0$, $\cdots$, $\rho_{r-1}\ne 0$, $\rho_r=0$. Then, as
$s\to\infty$, $G^{(\alpha)}$ tends to a diagonal matrix for
$\alpha\le r-1$. Considering (\ref{eq:0S}), the limit of 
$(G^{(r)}(\lambda))_{B_N}$ is also diagonal, hence
\begin{equation} 
   U^{[m]}=i[J,(G_1)_{B_N}]\to 0
\end{equation}

Now we turn to prove part (2) of the proposition. First, suppose
$r=2$. We use an alternate way of constructing Darboux
matrices.\zcite{bib:Zhou2n} Let
\begin{equation}
   \odotH_\alpha=\pmatrix \exp(Q_\alpha(s))\\ C_\alpha \endpmatrix\qquad
   \varGamma_{\alpha\beta}=\frac{\odotH_\alpha^*\odotH_\beta}
    {\lambda_\beta-\bar\lambda_\alpha}
\end{equation}
then
\begin{equation}
   G(\lambda)=\prod_{\alpha=1}^r(\lambda-\bar\lambda_\alpha)\left(
   1-\sum_{\alpha,\beta=1}^r
    \frac{\odotH_\alpha(\varGamma^{-1})_{\alpha\beta}\odotH_\beta^*}
    {\lambda-\bar\lambda_\beta}\right)
\end{equation}
and
\begin{equation}
   U^{[m]}=i\left[J,\sum_{\alpha,\beta=1}^r
    \left(\odotH_\alpha(\varGamma^{-1})_{\alpha\beta}
    \odotH_\beta^*\right)_{B_N}\right]
\end{equation}

Case (i): $l_\alpha=l_\beta$.

\begin{equation}
   \aligned
   &\odotH_\alpha\doteq\pmatrix 1 &&&&&&\\ &\ddots &&&&&\\ &&1 &&&&\\
     &&&\exp(\pi_{\alpha}) &&&\\ &&&&1 &&\\ 
     &&&&&\ddots &\\ &&&&&&1\\ 0 &\cdots &0
     &\underset{\scriptstyle l_\alpha}\to\kappa_\alpha &0 &\cdots &0
    \endpmatrix\\
   &\odotH_\beta\doteq\pmatrix 1 &&&&&&\\ &\ddots &&&&&\\ &&1 &&&&\\
     &&&\exp(\pi_{\beta}) &&&\\ &&&&1 &&\\ 
     &&&&&\ddots &\\ &&&&&&1\\ 0 &\cdots &0 
     &\underset{\scriptstyle l_\beta}\to\kappa_\beta &0 &\cdots &0
    \endpmatrix
   \endaligned \label{eq:HaHb}
\end{equation}
where $l_\beta=l_\alpha$, then $\D\varGamma=\pmatrix \varGamma_{11}
&\varGamma_{12}\\ \varGamma_{21} &\varGamma_{22}\endpmatrix$
where $\varGamma_{jk}$'s are $N\times N$ diagonal matrices.
Therefore, $\D\varGamma^{-1}=\pmatrix \varSigma_{11}
&\varSigma_{12}\\ \varSigma_{21} &\varSigma_{22}\endpmatrix$
where $\varSigma_{jk}$'s are also $N\times N$ diagonal matrices.
This implies that $U^{[m]}=0$. 

Case (ii): $l_\alpha\ne l_\beta$.
Suppose $\odotH_\alpha$, $\odotH_\beta$ are given by
(\ref{eq:HaHb}) with $l_\beta\ne l_\alpha$. Denote
\begin{equation}
   \theta_{\alpha\beta}=\frac{\kappa_\alpha^*\kappa_\beta}
    {\sqrt{\kappa_\alpha^*\kappa_\alpha\kappa_\beta^*\kappa_\beta}}
   \label{eq:deftheta}
\end{equation}
\begin{equation}
   g_{\alpha\beta}=1-\frac{4\im\lambda_\alpha\im\lambda_\beta}
    {|\lambda_\beta-\bar\lambda_\alpha|^2}|\theta_{\alpha\beta}|^2>0
   \label{eq:defg}
\end{equation}
then, by direct computation, we have
\begin{equation}
   \aligned
   &\lim_{s\to\infty}U_{l_\alpha,l_\beta}^{[m]}\exp(i(\phi_\beta-\phi_\alpha)s)
    =\frac{2i(J_{l_\beta}-J_{l_\alpha})\im\lambda_\alpha\im\lambda_\beta
    \theta_{\alpha\beta}}{\bar\lambda_\beta-\lambda_\alpha}\;
    \frac{\exp(i\im(\pi_{\alpha}-\pi_{\beta}))}{D}\\
   &D=\sqrt{g_{\alpha\beta}}\cosh(\re(\pi_{\alpha}+\pi_{\beta})-\delta_1)
    +\cosh(\re(\pi_{\alpha}-\pi_{\beta})-\delta_2)\\
   &\delta_1=\frac 12\ln g_{\alpha\beta}
    +\frac 12\ln(\kappa_\alpha^*\kappa_\alpha\kappa_\beta^*\kappa_\beta)
    +2\ln\left|\frac{\lambda_\beta-\lambda_\alpha}
    {\lambda_\beta-\bar\lambda_\alpha}\right|\\
   &\delta_2=\frac 12
    \ln\frac{\kappa_\alpha^*\kappa_\alpha}{\kappa_\beta^*\kappa_\beta}
   \endaligned
   \label{eq:stnexpr}
\end{equation}
and $U_{\mu\nu}\to 0$ if $(\mu,\nu)\ne (l_\alpha,l_\beta)$.

When $r>2$, we still use the permutability of Darboux
transformations and suppose $\rho_1\ne 0$, $\cdots$,
$\rho_{r-2}\ne 0$, $\rho_{r-1}=\rho_r=0$. Then, after the action
of $G^{(r-2)}(\lambda)\cdots G^{(1)}(\lambda)$, the derived
$H_{r-1}^{(r-2)}$, $H_r^{(r-2)}$ has the same asymptotic form as
$H_{r-1}$, $H_r$, provided that the constant vectors
$\kappa_{r-1}$, $\kappa_r$ are changed to $\kappa_{r-1}^{(r-2)}$,
$\kappa_r^{(r-2)}$. Therefore, as in the case $r=2$, the limit of
$U_{l_{r-1},l_r}$ has the desired form, and the limits of the
other components of $U$ are all zero. The proposition is proved. 
\end{demo}

\section{Localization of the solutions}\label{sec:loc}

Now we consider (\ref{eq:LP2}), For this system,
\begin{equation}
   Q_\alpha=i\lambda_\alpha(x+Jy)+i\omega(\lambda_\alpha)t
\end{equation}
We consider the limit of the solution as $(x,y)\to\infty$ along a
straight line $x=\xi+v_xs$, $y=\eta+v_ys$ ($v_x^2+v_y^2>0$), then
\begin{equation}
   Q_\alpha=i\lambda_\alpha(\xi+J\eta)+i\omega(\lambda_\alpha)t
   +i\lambda_\alpha(v_x+Jv_y)s
\end{equation}
Now
\begin{equation}
   \rho_\alpha=\re\left(i\lambda_\alpha(v_x+J_{l_\alpha}v_y)\right)
   =-\im\lambda_\alpha(v_x+J_{l_\alpha}v_y)
\end{equation}
If there is at most one $\rho_\alpha=0$, then
Proposition~\ref{theprop} implies that $U^{[m]}\to 0$ as
$s\to\infty$. If $\rho_\alpha=0$, $\rho_\beta=0$ $(\alpha\ne
\beta)$, then $l_\alpha=l_\beta$ since $J_{l_\alpha}\ne
J_{l_\beta}$ if $l_\alpha\ne l_\beta$. Hence,
Proposition~\ref{theprop} also implies $U^{[m]}\to 0$ as 
$s\to\infty$. Therefore, we have
\begin{theorem}\label{thm1}
$U^{[m]}\to 0$ as $(x,y)\to\infty$ in any directions.
\end{theorem}

\section{Asymptotic behavior of the solutions as $t\to\infty$}

Now we use a frame $(\xi,\eta)$ which moves in a fixed velocity
$(v_x,v_y)$, that is, let $x=\xi+v_xt$, $y=\eta+v_yt$, then
\begin{equation}
   Q_\alpha=i\lambda_\alpha(\xi+J\eta)
   +(i\lambda_\alpha(v_x+Jv_y)+i\omega(\lambda_\alpha))t
\end{equation}
\begin{equation}
   \rho_\alpha=-\im\lambda_\alpha(v_x+J_{l_\alpha}v_y)
    -\im(\omega_{l_\alpha}(\lambda_\alpha))
\end{equation}

Suppose that for mutually different $\alpha$, $\beta$, $\gamma$,
\begin{equation}
   \det\pmatrix 1 &J_{l_\alpha} &\sigma_{\alpha}\\
   1 &J_{l_\beta} &\sigma_{\beta}\\ 1 &J_{l_\gamma} &\sigma_{\gamma} 
   \endpmatrix\ne 0
   \label{eq:diffv}
\end{equation}
where 
\begin{equation}
   \sigma_\alpha=\im(\omega_{l_\alpha}(\lambda_\alpha))/\im(\lambda_\alpha)
   \label{eq:defsigma}
\end{equation}
Then there are at most two $\rho_\alpha=0$ $(\alpha=1,\cdots,r)$. By
Proposition~\ref{theprop}, $U^{[m]}_{l_\alpha,l_\beta}\not\to 0$ only if
$\rho_\alpha=0$, $\rho_\beta=0$. This leads to
\begin{equation}
   v_x=\frac{J_{l_\alpha}\sigma_\beta-J_{l_\beta}\sigma_\alpha}
    {J_{l_\beta}-J_{l_\alpha}}\qquad
   v_y=\frac{\sigma_\alpha-\sigma_\beta}{J_{l_\beta}-J_{l_\alpha}}
   \label{eq:velo}
\end{equation}

For $U_{jk}^{[m]}\not\to 0$, $\alpha$, $\beta$ can take $m_j$ and
$m_k$ values respectively, hence there are at most $m_jm_k$
velocities $(v_x,v_y)$ such that $U_{jk}^{[m]}\not\to 0$.
Therefore, we have 
\begin{theorem}\label{thm2}
Suppose (\ref{eq:diffv}) is satisfied. Then as $t\to\infty$, the
asymptotic solution of $U_{jk}^{[m]}$ has at most $m_jm_k$ peaks
whose velocities are given by (\ref{eq:velo}) ($l_\alpha=j$,
$l_\beta=k$). If a peak has velocity $(v_x,v_y)$, then, in the
coordinate $\xi=x-v_xt$, $\eta=y-v_yt$,
$\lim_{t\to\infty}U_{ab}=0$ for all $(a,b)\ne (j,k)$, and as $t\to\infty$
\begin{equation}
   \aligned
   &U_{jk}^{[m]}\sim
    \frac{B_{\alpha\beta}\exp(i\re(\lambda_\alpha-\lambda_\beta)\xi
    +i(\lambda_\alpha J_j-\lambda_\beta J_k)\eta
    +i(\phi_\alpha-\phi_\beta)t)}D\\
   &D=A_{\alpha\beta}\cosh(\im(\lambda_\alpha+\lambda_\beta)\xi
    +\im(\lambda_\alpha J_j+\lambda_\beta J_k)\eta
    +\delta_{\alpha\beta}^{(1)})\\
   &\quad+\cosh(\im(\lambda_\alpha-\lambda_\beta)\xi
    +\im(\lambda_\alpha J_j-\lambda_\beta J_k)\eta
    +\delta_{\alpha\beta}^{(2)}) 
   \endaligned
\end{equation}
where $A_{\alpha\beta}$, $\delta_{\alpha\beta}^{(1)}$,
$\delta_{\alpha\beta}^{(2)}$ are real constants,
$A_{\alpha\beta}>0$, and $B_{\alpha\beta}$ is a complex constant,
\begin{equation}
   \phi_\gamma=\re\lambda_\gamma(v_x+J_{l_\gamma}v_y)
   +\re(\omega_{l_\gamma}(\lambda_\gamma))\quad 
   (\gamma=\alpha,\beta)
\end{equation}
\end{theorem}

{\sl Remark:}
The condition (\ref{eq:diffv}) implies that the velocities of the
solitons are all different. This is true for the DSI equation.
However, for the N-wave equation, all the solitons move in the
same velocity. We will discuss this problem in the next section.

{\bf Example:} DSI equation

Let $n=2$, $N=2$,
\begin{equation}
   J=\pmatrix 1 &0\\ 0 &-1 \endpmatrix\qquad
   U=\pmatrix 0 &u\\ -\bar u &0\endpmatrix\qquad
   \omega=-2iJ\lambda^2
\end{equation}
then we have
\begin{equation}
   \aligned
   &F_y=JF_x+UF\\
   &F_t=2iJF_{xx}+2iUF_x+i\pmatrix
    |u|^2+2q_1 &u_x+u_y\\ -\bar u_x+\bar u_y &-|u|^2-2q_2 \endpmatrix F
   \endaligned
   \label{eq:DSLP}
\end{equation}
\begin{equation}
   \aligned
   &-iu_t=u_{xx}+u_{yy}+2|u|^2u+2(q_1+q_2)u\\
   &q_{1,x}-q_{1,y}=q_{2,x}+q_{2,y}=-(|u|^2)_x
   \endaligned \label{eq:DS}
\end{equation}
\begin{equation}
   (FF^*)^D=\frac 12\pmatrix q_1 &0\\ 0 &q_2\endpmatrix,\qquad
   [J,FF^*]=U_x
   \label{eq:DSconstr}
\end{equation}

(\ref{eq:DS}) is the DSI equation.

If we construct the solution $U^{[m]}$ as above, then
Theorem~\ref{thm1} implies that $U^{[m]}\to 0$ as
$(x,y)\to\infty$ in any directions. If
$\re\lambda_\alpha\ne\re\lambda_\beta$ for $\alpha\ne\beta$ and 
$l_\alpha=l_\beta$, then, theorem~\ref{thm2} implies that as
$t\to\infty$, $u$ has at most $m_1m_2$ peaks ($m_1+m_2=r$). From
(\ref{eq:defsigma}),
$\sigma_\alpha=-4J_{l_\alpha}\re\lambda_\alpha$, hence 
(\ref{eq:velo}) implies that each peak has the velocity
$v_x=2\re(\lambda_\alpha-\lambda_\beta)$,
$v_y=2\re(\lambda_\alpha+\lambda_\beta)$ 
$(l_\alpha=1,\,l_\beta=2)$. 
This is the $(m_1,m_2)$ solitons.\zcite{bib:FS}
When $K=1$, these peaks do not vanish if and only if
$\kappa_\alpha$'s are all nonzero.

Fig.~1 -- 3 shows the solitons $u^{[1,3]}$, $u^{[2,3]}$ and
$u^{[3,3]}$ respectively. The parameters are $K=1$, $t=2$,
$\lambda_1=1-2i$, $\lambda_2=-3-i$, $\lambda_3=2+i$, 
$\lambda_4=-1+3i$, $\lambda_5=2+1.5i$, $\lambda_6=-0.5-1.5i$,
$C_1=(1,0)$, $C_2=(0,1)$, $C_3=(0,2)$, $C_4=(0,-2)$, 
$C_5=(2,0)$, $C_6=(-2,0)$. Some techniques on drawing such
figures were explained in \cite{bib:Zhoufig}.

\unitlength=1mm
\vbox{%
\ifx\figuretype\BMPfile
{
\begin{picture}(60,32)
\put(16,-15){\special{em:graph Zhou01.bmp}}
\end{picture}
\vskip1.5cm
}
\else
{
\vskip-2cm
\epsffile[-60 0 400 222]{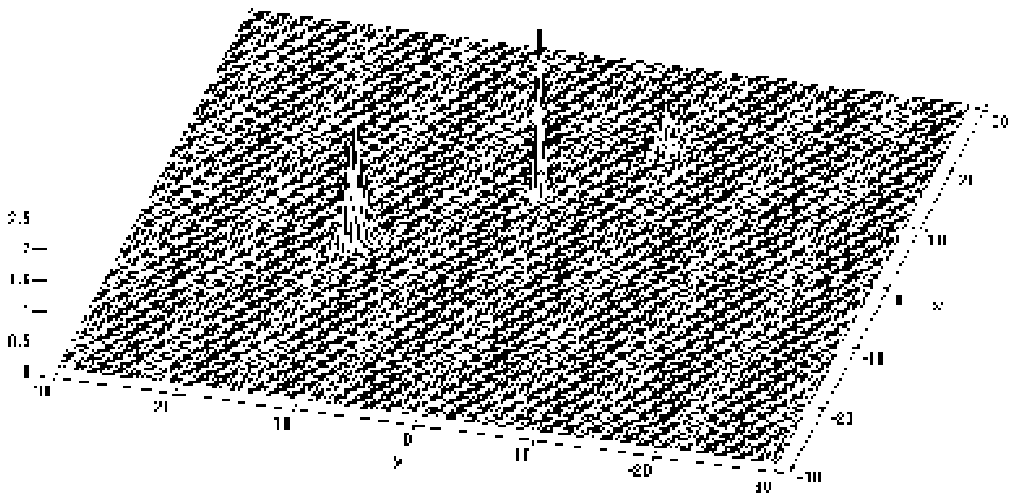}
}
\fi
\hbox to \hsize{\hfill\scriptsize Fig.~1. $u^{[1,3]}$ of 
the DSI equation\hfill}}
\vskip1cm

\vbox{%
\ifx\figuretype\BMPfile
{
\begin{picture}(60,32)
\put(16,-15){\special{em:graph Zhou02.bmp}}
\end{picture}
\vskip1.5cm
}
\else
{
\vskip-2cm
\epsffile[-60 0 400 222]{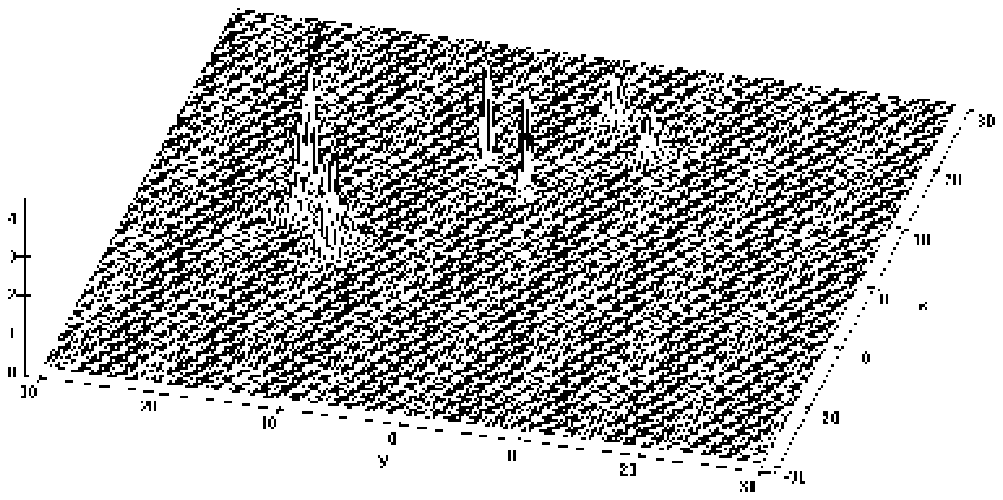}
}
\fi
\hbox to \hsize{\hfill\scriptsize Fig.~2. $u^{[2,3]}$ of 
the DSI equation\hfill}}
\vskip1cm

\vbox{%
\ifx\figuretype\BMPfile
{
\begin{picture}(60,32)
\put(16,-15){\special{em:graph Zhou03.bmp}}
\end{picture}
\vskip1.5cm
}
\else
{
\vskip-2cm
\epsffile[-60 0 400 222]{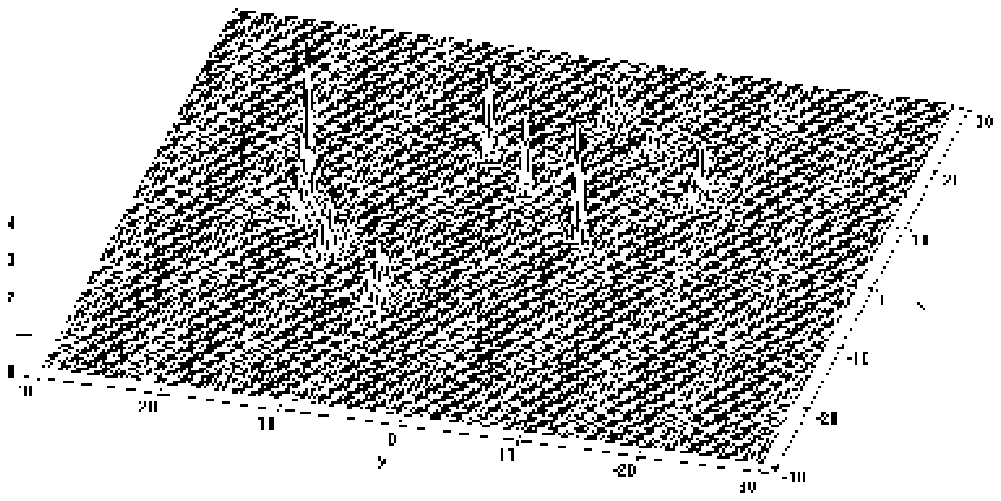}
}
\fi
\hbox to \hsize{\hfill\scriptsize Fig.~3. $u^{[3,3]}$ of 
the DSI equation\hfill}}
\vskip1cm

\section{Asymptotic solutions as the phases differences tend to
infinity} 

For the equations whose solitons move in the same speed, like the
N-wave equation, the peaks do not separate as $t\to\infty$.
However, we can still see some peaks in the
figures.\zcite{bib:Zhou3wave}  In \cite{bib:Zhou3wave}, we
obtained the asymptotic behavior of the $r^2$ solitons. Here we
will get the corresponding asymptotic properties of more general
solitons obtained in this paper. We have
\begin{theorem}\label{thm3}
Let $p_\alpha$ $(\alpha=1.\cdots,r)$ be constant real numbers
satisfying 
\begin{equation}
   \det\pmatrix 1 & J_{l_\alpha} &p_\alpha/\im\lambda_\alpha\\
    1 & J_{l_\beta} &p_\beta/\im\lambda_\beta\\
    1 & J_{l_\gamma} &p_\gamma/\im\lambda_\gamma\\ \endpmatrix\ne 0
   \label{eq:Ndiffv}
\end{equation}
for mutually different $\alpha$, $\beta$, $\gamma$. Let $d_\alpha$
be complex constant $K\times 1$ vectors,
$\kappa_\alpha=d_\alpha\exp(p_\alpha\tau)$ and construct the
Darboux transformations as above. Let $x=\xi+v_x\tau$,
$y=\eta+v_y\tau$, then, for any $j$, $k$ with $1\le j,k\le N$,
$j\ne k$, $\lim_{\tau\to\infty} U_{jk}^{[m]}\ne 0$ only if
$(v_x,v_y)$ takes specific $m_jm_k$ values. 
\end{theorem}

\begin{demo}
As in \S\ref{sec:loc}, here
\begin{equation}
   Q_\alpha=i\lambda_\alpha(\xi+J\eta)+i\omega(\lambda_\alpha)t
   +i\lambda_\alpha(v_x+Jv_y)\tau,
\end{equation}
hence
\begin{equation}
   \odotH_\alpha\doteq\pmatrix \exp(\widetilde Q_\alpha(\tau))\\
    D_\alpha\endpmatrix
\end{equation}
where
\begin{equation}
   D_\alpha=(0,\cdots,0,\underset{l_\alpha}\to{d_\alpha},0,\cdots,0)
\end{equation}
\begin{equation}
   \widetilde Q_\alpha(\tau)=i\lambda_\alpha(\xi+J\eta)
   +i\omega(\lambda_\alpha)t+(i\lambda_\alpha(v_x+Jv_y)-p_\alpha)\tau,
\end{equation}
hence the real part of the coefficient of $\tau$ in $\widetilde
Q_\alpha(\tau)$ is 
\begin{equation}
   \widetilde\rho_\alpha=-\im\lambda_\alpha(v_x+Jv_y)-p_\alpha
\end{equation}
Condition (\ref{eq:Ndiffv}) implies that there are at most two
$\widetilde\rho_\alpha$'s such that $\widetilde\rho_\alpha=0$. According to
Proposition~\ref{theprop}, as $\tau\to\infty$,
$U_{jk}^{[m]}\not\to 0$ only if there exist $\widetilde\rho_\alpha=0$,
$\widetilde\rho_\beta=0$, $\alpha\ne \beta$, $l_\alpha=j$, $l_\beta=k$. 
Therefore, the theorem is verified.
\end{demo}

When the condition (\ref{eq:diffv}) holds, this theorem is
useless, because the evolution will always separate the peaks,
However, when (\ref{eq:diffv}) does not hold, especially when it
is never satisfied, this theorem reveals a fact of the separation
of the peaks.

{\bf Example:} N-wave equation

Let $n=1$, $\omega=L\lambda$ where $L=\diag(L_1,\cdots,L_N)$ is
a constant real diagonal matrix such that $L_j\ne L_k$ for $j\ne
k$. Then, the integrability conditions (\ref{eq:redLP}) --
(\ref{eq:redevol}) imply 
\begin{eqnarray}
   &&F_y=JF_x+UF\qquad F_t=LF_x+VF \\
   &&[J,V]=[L,U]\qquad U_t-V_y+[U.V]+JV_x-LU_x=0 \label{eq:Nwave}\\
   &&U_x=[J,FF^*]
\end{eqnarray} 
(\ref{eq:Nwave}) is just the N-wave equation.

Suppose $U^{[m]}$ is constructed as above, then Theorem~\ref{thm1}
implies that $U^{[m]}\to 0$ as $(x,y)\to\infty$ in any directions.
Theorem~\ref{thm2} cannot be applied here. The reason is: the
condition (\ref{eq:diffv}) holds only if $l_\alpha\ne l_\beta$
for $\alpha\ne \beta$. Hence for any $j$, $m_j=0$ or $1$. This
implies that (\ref{eq:diffv}) does not hold generally unless
$m_j=0$ or $1$ for all $1\le j\le N$. Therefore, we apply
Theorem~\ref{thm3} to the previous problem. Theorem~\ref{thm3}
implies that if we choose $\{p_\alpha\}$ such that
(\ref{eq:Ndiffv}) is satisfied, then, for each $(j,k)$,
$\lim_{\tau\to\infty}U_{jk}^{[m]}$ has at most $m_jm_k$ peaks. 
When $K=1$, these peaks do not vanish if and only if
$\kappa_\alpha$'s are all nonzero.

{\sl Remark:}
Here $\tau\to\infty$ means that the phase differences of
different peaks tend to infinity. Therefore, the peaks are
separated by enlarging the phase differences.

Here are the figures describing the solutions $U^{[0,1,2]}$ and
$U^{[1,1,2]}$ of the 3-wave equation. The vertical axis is
$(|u_{12}|^2+|u_{13}|^2+|u_{23}|^2)^{1/4}$ so that all the
components are shown in one figure. The parameters are 
$$ J=\pmatrix 1&&\\ &0&\\ &&-1 \endpmatrix\qquad
   L=\pmatrix 2&&\\ &-1&\\ &&1 \endpmatrix
$$
$K=1$, $t=10$, $\lambda_1=1-2i$, $\lambda_2=-3-i$,
$\lambda_3=2+i$, $\lambda_4=-1+3i$, $C_1=(0,1,0)$,
$C_2=(0,0,1)$, $C_3=(0,0,4096)$, $C_4=(1,0,0)$.
Note that for $U^{[0,1,2]}$, only $U_{23}$ has two peaks, and for
$U^{[1,1,2]}$, $U_{12}$, $U_{13}$, $U_{23}$ have one, two, two
peaks respectively.

\vbox{%
\ifx\figuretype\BMPfile
{
\begin{picture}(60,32)
\put(16,-15){\special{em:graph Zhou04.bmp}}
\end{picture}
\vskip1.5cm
}
\else
{
\vskip-2cm
\epsffile[-60 0 400 222]{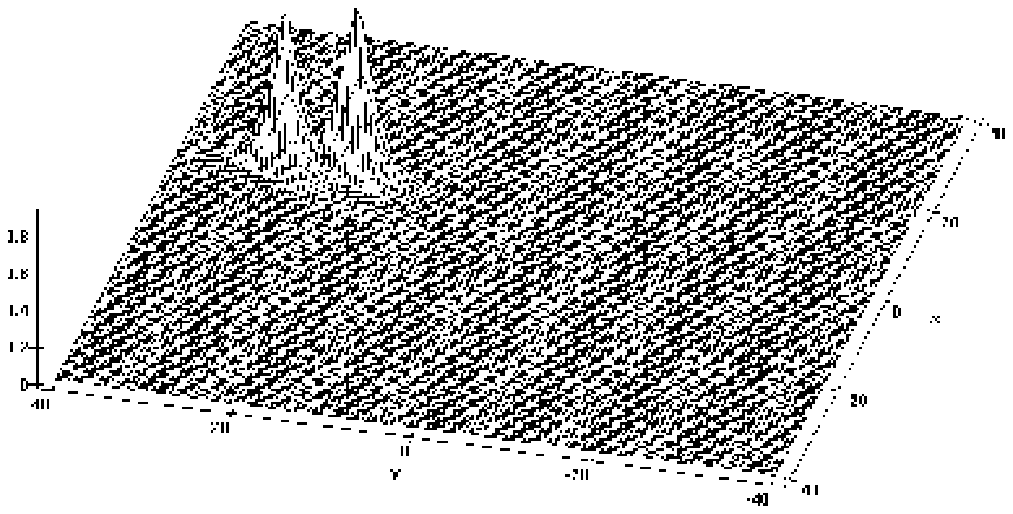}
}
\fi
\hbox to \hsize{\hfill\scriptsize Fig.~4. $U^{[0,1,2]}$ of 
the 3-wave equation\hfill}}
\vskip1cm

\vbox{%
\ifx\figuretype\BMPfile
{
\begin{picture}(60,32)
\put(16,-15){\special{em:graph Zhou05.bmp}}
\end{picture}
\vskip1.5cm
}
\else
{
\vskip-2cm
\epsffile[-60 0 400 222]{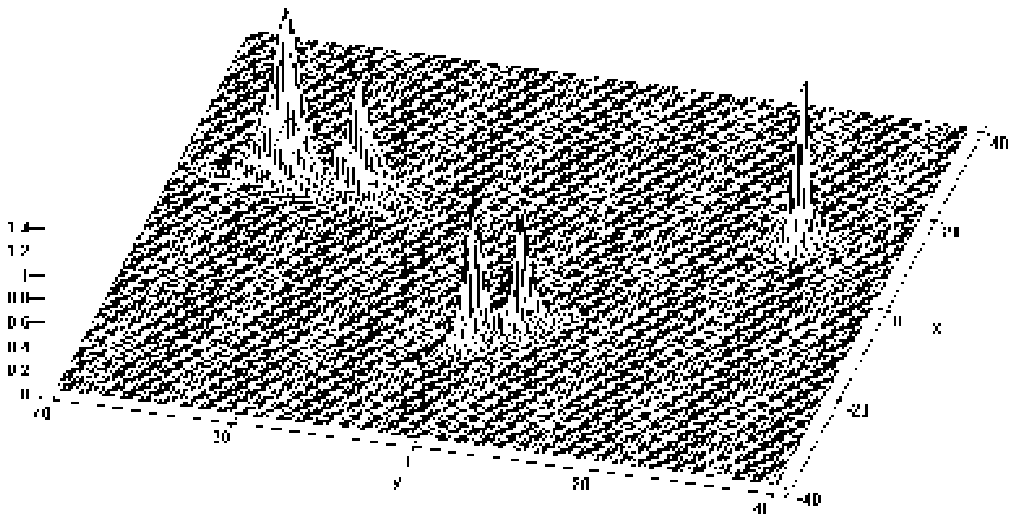}
}
\fi
\hbox to \hsize{\hfill\scriptsize Fig.~5. $U^{[1,1,2]}$ of 
the 3-wave equation\hfill}}

\end{document}